\documentclass[preprint,authoryear,3p]{elsarticle}
\usepackage{amsmath}
\usepackage{amsfonts}
\usepackage{amssymb}
\usepackage{theorem}
\usepackage{graphicx}
\usepackage[cp1250]{inputenc}

\usepackage[english]{babel}
\usepackage{hyperref}

\usepackage{color}

\newtheorem{remark}{Remark}[section]

\begin{document}
\begin{frontmatter}
\title{Pricing electricity derivatives within a Markov regime-switching model}

\author{Joanna Janczura}
\ead{joanna.janczura@pwr.wroc.pl}
\address{Hugo Steinhaus Center, Institute of Mathematics and Computer Science,\\ 
Wroc{\l}aw University of Technology, 50-370 Wroc{\l}aw, Poland}

\date{This version: \today}

\begin{abstract}
In this paper analytic formulas for electricity derivatives are calculated. To this end, we assume that electricity spot prices follow a 3-regime Markov regime-switching model with independent spikes and drops and periodic transition matrix. Since the classical derivatives pricing methodology cannot be used in case of non-storable commodities, we employ the concept of the risk premium. The obtained theoretical results are then used for the European Energy Exchange (EEX) market data. The 3-regime model is calibrated to the spot electricity prices. Next, the risk premium is derived and used to calculate prices of European options written on spot, as well as, forward prices. 
\end{abstract}


\end{frontmatter}

\section{Introduction}
Deregulation of electricity markets has led to a substantial increase in risk borne by market participants. The often unexpected, extreme spot price changes range even two orders of magnitude and can cause severe financial problems to the utilities that buy electricity in the wholesale market and deliver it to consumers at fixed prices. The utilities and other power market companies need to hedge against this price risk. A straightforward way to do it is to use derivatives, like forwards and options. It is exactly the aim of this paper to price commonly traded electricity derivatives.

Before calculating a price of a derivative, a proper model for the underlying asset has to be chosen. There are two approaches common for the electricity market. The first one is to start with specifying the forward price dynamics \citep[see e.g.\ ][]{cle:str:99, ben:koe:08, bje:etal:10}. Such approach is useful if only derivatives written on forwards are to be considered and a link between forward and spot prices is not important for modeling issues. Obviously, the spot price can always be derived using the fact that the forward and spot prices should coincide at the forward settlement. However, the complexity of the spot price dynamics is then usually neglected.
The second approach is based on defining the spot price dynamics first \citep[see e.g.\ ][]{luc:sch:02,mil:03, ben:eke:hau:nie:03, bie:etal:07} and then modeling a link between the forward and spot markets. Usually, a convenience yield or risk premium notion is used \citep{ben:ben:koe:08,gem:05,wer:06}. Using such an approach allows to price derivatives written both on the spot and the forward price. Moreover, a relation between spot and forward prices is taken into account and the lack of forward price data is no more a limitation. Here, we use the latter approach and describe the spot price dynamic by a Markov regime-switching (MRS) model with independent spikes and drops and periodic transition matrix that was proposed by \cite{jan:wer:10:ee}. 

After specifying a model we have to choose derivatives pricing methodology. Classical approach used in financial and commodities markets is based on the no-arbitrage assumption and construction of a strategy replicating a future payoff (or equivalently finding a martingale measure, \cite{mus:rut:97}). 
However, such approach fails in case of electricity due to very limited storage possibilities. Therefore, instead of using a martingale approach, we employ a concept of the risk premium/market price of risk and find such pricing measure that yields the observed forward market prices. With such methodology we are able to derive forward prices from the spot price model and also to find explicit formulas for premiums of European options written on spot, as well as, on forward prices. 

The paper is structured as follows. In Section \ref{sec:model} we introduce the Markov regime-switching model used for electricity spot price dynamics. Next, in Section \ref{sec:rp} we explain why the classical derivatives pricing approach fails in case of electricity derivatives and, as a solution, we describe the `risk premium' approach in case of the considered model. The obtained results are then used to derive analytical formulas for prices of electricity derivatives in Section \ref{sec:pr}. Finally, in Section \ref{sec:ex} we use the obtained theoretical results to price derivatives using the European Energy Exchange (EEX) data and in Section \ref{sec:con} we conclude.

\section{The model}\label{sec:model}

Let us first recall the main stylized facts about electricity prices. Electricity prices highly depend on the actual demand/consumption. Obviously, the latter is varying during a year due to the changing weather conditions and throughout the week or day due to the business cycle. The same long-term (yearly) and short-term (weekly/daily) seasonality is recorded for electricity prices. The second apparent feature is the very high volatility of electricity prices and unexpected, usually transient, dramatic price changes called spikes or jumps. The spot price may rise for a few hours to even two orders of magnitude of the standard prices and then fall back to the normal level. What makes electricity market completely different from other financial or commodities markets is that the electricity prices may, as well, abruptly fall down yielding negative values. Finally, electricity prices are mean-reverting, meaning that in long time period they move back to some equilibrium level.

Seasonality is usually removed from the analyzed prices prior to modeling by fitting some periodic function like sine \citep{pil:98, car:fig:05, dej:06} or piecewise constants \citep{luc:sch:02, kni:rob:05}. Alternatively, some smoothing technique like wavelets or moving average can be used \citep{wer:09:mmor}. The mean-reverting property is typically modeled with some mean-reverting processes, like e.g.\ AR(1) time series or the \cite{vas:77} model. The most challenging for modeling, and at the same time the most important for risk management, are the price spikes. One approach is to incorporate a jump component into a standard mean-reverting diffusion model \citep{den:98, car:fig:05, wer:08}. However, in the resulting jump-diffusion models there is a problem of how the price after a spike revert back to the normal level. Nor an immediate negative jump, nor mean-reversion pulling back the prices to the normal level, yields a flexible tool for modeling consecutive spikes. 
Another possibility are the Markov regime-switching (MRS) models in which the prices might stay in the excited (spike) regime with some probability. Hence, MRS models allow for modeling consecutive spikes in a very natural way and seem to be a reasonable choice for electricity price dynamics. To our best knowledge MRS models were first applied to electricity prices by \cite{eth:mou:98} who used an AR time series with parameters depending on the actual regime. A MRS model with independent spikes was later introduced by \cite{hui:dej:03} and \cite{dej:06}. Numerous attempts improving statistical properties of the model \citep[see e.g.\ ][]{hui:mah:03,mou:nin:cai:06,hal:nie:06,wer:09:mmor} or including some exogenous factors \citep[see e.g.\ ][]{hui:08, kan:oha:08, kar:bun:08} were later proposed. Here, we focus on a 3-regime model with independent spikes and drops introduced recently by \cite{jan:wer:10:ee}.

Having in mind the above mentioned features of electricity, we let the electricity spot price be given by
\begin{equation}\label{eqn:pr:P}
P_t=g_t+X_t,
\end{equation}
where $g_t$ is a deterministic seasonal component and $X_t$ follows the 3-regime model with independent spikes and drops. Namely,
\begin{equation}
X_t=\left\{\begin{array}{lcr} 
X_{t,b}& \mbox{ if } &R_{\lfloor t\rfloor}=b, \\
X_{\lfloor t\rfloor,s}& \mbox{ if } &R_{\lfloor t\rfloor}=s, \\
X_{\lfloor t\rfloor,d}& \mbox{ if } &R_{\lfloor t\rfloor}=d,
 \end{array}\right.\label{eqn:model_def}\end{equation}
where $b$ denotes the base regime (describing the `normal' prices), $s$ the spike regime (representing the sudden upward price jumps), while $d$ stands for the drop regime (responsible for the sudden price drops). Further, $\lfloor t\rfloor$ denotes the integer part of $t$, $R_k$ is a discrete-time Markov chain defined by a time-varying (periodic) transition matrix $\mathbf{P}(k)$ 
\begin{equation}\label{eqn:MRS:transition_matrix}
\mathbf{P}(k)=(p_{ij}(k))=
\left(
\begin{array}{ccc}
 p_{bb}(k) & p_{bs}(k) & p_{bd}(k)\\
 p_{sb}(k) & p_{ss}(k) & p_{sd}(k)\\
 p_{db}(k) & p_{ds}(k) & p_{dd}(k)\\
\end{array}
\right),
\end{equation}
for $k\in \{0,1,2,...\}$. The base regime dynamics is given by the \cite{vas:77} model:
\begin{equation}\label{eqn:pr:OU}
dX_{t,b}=(\alpha-\beta X_{t,b})dt+\sigma_bdW_t,
\end{equation} 
having unique mean-reverting solution of the form:
\begin{equation}\label{eqn:mr:OU:sol}
X_{t}=X_{0}e^{-\beta t}+\frac{\alpha}{\beta}\left(1-e^{-\beta t}\right)+\sigma\int_0^te^{-\beta (t-u)}dW_u,
\end{equation}
where $W_t$ is a Wiener process (or Brownian motion), $\beta$ is the speed of mean-reversion, $\frac{\alpha}{\beta}$ is the long-time equilibrium level and $\sigma$ is the volatility.
The spike regime values $(X_{0,s},X_{1,s},X_{2,s},...)$ constitute an i.i.d.\ sample from the $c_s$-shifted log-normal distribution, i.e.\ :
\begin{equation}\label{eqn:pr:SLN}
\ln(X_{k,s}-c_s)\sim N(\mu_s,\sigma_s^2), \quad X_{k,s}>c_s
\end{equation}
while the drop regime values $(X_{0,d},X_{1,d},X_{2,d},...)$ form an i.i.d.\ sample from the inverted $c_d$-shifted log-normal distribution defined as:
\begin{equation}\label{eqn:pr:ISLN}
\ln(-X_{k,d}+c_d)\sim N(\mu_d,\sigma_d^2), \quad X_{k,d}<c_d.
\end{equation}

Observe, that in the model defined by (\ref{eqn:model_def}) the price process can jump to a different regime only at discrete time points $t=0,1,2,...$. This is motivated by the fact, that, even though the price is a result of continuous bidding, electricity spot price is typically settled for contracts with some delivery period, usually an hour. Hence, the change in electricity spot price dynamics may occur only in discrete time points.
 
Finally, let $(\Omega,\mathcal{F},Q)$ be a probability space with filtration $\mathcal{F}_t$ generated by the processes $X_t$ and $R_{\lfloor t\rfloor}$ and assume a constant continuously compounded interest rate $r$. 

\section{The risk premium}\label{sec:rp}
The classical option pricing approach is based on the no-arbitrage assumption, which implies that the fair price of a derivative is the discounted expected future payoff under a martingale measure \citep{har:pli:83}. If the market is complete, i.e.\ any contingent claim can be replicated with a self-financing strategy (it is \textit{attainable}), there exists a unique martingale measure. However, due to the non-storability of electricity a derivative written on the spot electricity price cannot be replicated with a portfolio consisting of the underlying instrument and a financing debt account. As a consequence, the market is incomplete. Moreover, since electricity cannot be traded in the usual way (once purchased has to be consumed), the only tradable asset in the spot market is the bank account \citep{ben:eke:hau:nie:03}. Recall that here we assume a continuously compounded constant interest rate $r$. Obviously, the discounted value of the bank account is a martingale under any measure equivalent to the actual (also called the objective or statistical) measure $Q$. Therefore, the market is arbitrage-free but there is no unique martingale measure. An additional criterion has to be used in order to select a pricing measure. 

Here we use an approach based on the concept of the risk premium \citep[see e.g.\ ][]{ben:ben:koe:08,gem:05,wer:06}, which is defined as a reward for investing into a risky asset instead of a risk-free one. Note, that a related notion is that of the market price of risk, which can be seen as a drift adjustment in the dynamics of an asset to reflect how investors are compensated for bearing risk when holding the asset \citep{ben:car:kie:08}. The idea of the `risk premium' approach is to choose a martingale measure that is consistent with the prices of forward contracts quoted in the market. A similar approach is used in the weather \citep{ben:ben:07} or interest rate derivatives context and is based on calibrating the model to the initial yield curve \citep[see e.g.\ ][]{hul:whi:93, bjo:97}. 

Recall, that the arbitrage-free price of a forward contract should be equal to the expected future spot price under a pricing measure, namely
\begin{equation}\label{eqn:pr:mpr}
f_{0}^t=E^{\lambda}(P_t|\mathcal{F}_0),
\end{equation}
where $f_{0}^t$ is the price at time 0 of a forward contract with a delivery at time $t$, $P_t$ is the electricity spot price and $E^{\lambda}(\cdot)$ is the expected value with respect to the pricing probability measure $Q^{\lambda}$, equivalent to the actual measure $Q$. In an incomplete market relation (\ref{eqn:pr:mpr}) does not yield a unique forward price, as it is dependent on the choice of $Q^{\lambda}$. Here we choose $Q^{\lambda}$ such that relation (\ref{eqn:pr:mpr}) is consistent with market data, i.e.\ $E^{\lambda}(P_t|\mathcal{F}_0)$ is calibrated to the quotations of forward contracts. In other words, we choose the pricing measure that is used by the market. In the following we will assume that the measure $Q^\lambda$ is the probability measure under which the drift of the base regime process is parametrized by a function $\lambda(t)$ (i.e.\ the market price of risk) chosen so that $E^{\lambda}(P_t|\mathcal{F}_0)$ yields the market forward price $f_0^t$.
 
Before we find the measure $Q^{\lambda}$, we give a brief explanation of how to calculate the risk premium in the 3-regime model (\ref{eqn:pr:P})-(\ref{eqn:pr:ISLN}). Assume, that the forward price $f_{0}^t$ is given for any maturity $t$ and the spot price model parameters $\theta=(\alpha,\beta,\sigma_b,\mu_s,\sigma_s,c_s,\mu_d,\sigma_d,c_d,\mathbf{P})$ are known.

The risk premium $RP$ is defined as 
\begin{equation}\label{eqn:pr:rp}
RP(t)=E(P_t|\mathcal{F}_0)-f_{0}^t,
\end{equation}
where $f_0^t$ is the market price at time 0 of a forward contract with delivery at time $t$.
\begin{remark}
It should be noted that some authors \citep[see e.g.\ ][]{eyd:wol:03} define the risk premium as the difference between the forward price and the expected spot price, i.e.\ $-RP(t)$.
\end{remark}

Let  $p_{ij}^{(t)}=P(R_{\lfloor t\rfloor}=j|R_0=i)$ denote the probability of switching from state $i$ at time 0 to state $j$ at time $\lfloor t\rfloor$. For a constant probability matrix $\mathbf{P}$ it is given by the $ij$th element of the $\lfloor t\rfloor$th power of the transition matrix, i.e.\ $p_{ij}^{(t)}=\left(\mathbf{P}^{\lfloor t\rfloor}\right)_{ij}$. For a time-varying probability matrix it is given by $p_{ij}^{(t)}=\left(\prod_{k=0}^{\lfloor t\rfloor}\mathbf{P}(k)\right)_{ij}$.

In order to simplify the derivation, in the following we assume that $P(R_0=b)=1$ or equivalently $X_{0}=X_{0,b}$ a.s., i.e.\ at time 0 the process $X_t$ is in the base regime with probability 1. 

Now, we can derive a formula for the risk premium. Observe that
\begin{eqnarray*}
E(X_t|\mathcal{F}_0)&=&P(R_{\lfloor t\rfloor}=b|R_0=b)E(X_{t,b}|\mathcal{F}_0)+P(R_{\lfloor t\rfloor}=s|R_0=b)E(X_{\lfloor t\rfloor,s}|\mathcal{F}_0)+\\
\\&&+P(R_{\lfloor t\rfloor}=d|R_0=b)E(X_{\lfloor t\rfloor,d}|\mathcal{F}_0).
\end{eqnarray*}
Recall, that $X_{\lfloor t\rfloor,s}$ and $X_{\lfloor t\rfloor,d}$ are random variables independent of $\mathcal{F}_0$. Hence, $E(X_{\lfloor t\rfloor,j}|\mathcal{F}_0)=E(X_{\lfloor t\rfloor,j})$ for $j=s,d$. Moreover, 
from the assumption of $X_{0}=X_{0,b}$ we have that 
$E(X_{t,b}|\mathcal{F}_0)=E(X_{t,b}|X_{0,b})$. Hence
\begin{equation}\label{eqn:pr:exp}
E(X_t|\mathcal{F}_0)=p_{bb}^{(t)}E(X_{t,b}|X_{0,b})+p_{bs}^{(t)}E(X_{\lfloor t\rfloor,s})
+p_{bd}^{(t)}E(X_{\lfloor t\rfloor,d}).
\end{equation}
As a consequence, from (\ref{eqn:pr:P}), (\ref{eqn:mr:OU:sol}) and (\ref{eqn:pr:exp}), the risk premium in the 3-regime MRS model defined by equations (\ref{eqn:model_def})-(\ref{eqn:pr:ISLN}) is given by:
\begin{equation*}\label{eqn:rp}
RP(t)=p_{bb}^{(t)}\left[x_{0}e^{-\beta t}+\frac{\alpha}{\beta}\left(1-e^{-\beta t}\right)\right]+p_{bs}^{(t)}\left(e^{\mu_s+\frac{1}{2}\sigma_s^2}+c_s\right) +p_{bd}^{(t)}\left(-e^{\mu_d+\frac{1}{2}\sigma_d^2}+c_d\right)+g_t-f_{0}^t,
\end{equation*}
where $x_0$ is the stochastic part of the price observed at time $0$ and $f_{0}^t$ is the market forward price.

\begin{remark}\label{rem}
Observe that, if assumption that $X_{0}=X_{0,b}$ is not satisfied we have:
\begin{equation}\label{eqn:pr:E:hist}
E(X_{t,b}|\mathcal{F}_0)=\mathbb{I}_{\{R_0=b\}}E(X_{t,b}|X_{0,b})+\sum_{k=1}^{\infty}\mathbb{I}_{\{R_0\neq b, R_{-1}\neq b, R_{-2}\neq b,..., R_{-k+1}\neq b, R_{-k}=b\}}E(X_{t,b}|X_{-k+1,b}),
\end{equation}
where a negative time index is used for the historical (i.e.\ before the moment of valuation $t=0$) values of the process and for $u<t$ $E(X_{t,b}|X_{u,b})=X_{u,b}e^{-\beta (t-u)}+\frac{\alpha}{\beta}\left(1-e^{-\beta ( t-u)}\right)$.
Note, that formula (\ref{eqn:pr:E:hist}) is a consequence of the fact that the base regime values become latent if a spike or drop occurs.
Moreover,
\begin{equation*}
P(R_{\lfloor t\rfloor}=j|\mathcal{F}_0)=\sum_{i\in \{b,s,d\}}\mathbb{I}_{\{R_0=i\}}P(R_{\lfloor t\rfloor}=j|R_0=i)=
\sum_{i\in \{b,s,d\}}\mathbb{I}_{\{R_0=i\}}p_{ij}^{(t)}.
\end{equation*}
Thus, the risk premium calculation and all of the following results can be generalized to the case $R_{0}\neq b$.
\end{remark}

\section{Electricity derivatives pricing}\label{sec:pr}
\subsection{Options written on the electricity spot price}\label{sec:pr:spot}
Now, we turn to pricing of a European call option written on the electricity spot price. Recall, that the European option is a contract that gives the buyer the right to buy/sell the underlying commodity at some future date $t$ (called maturity) at a certain price $K$ (called the strike price). 
First, we find the pricing measure $Q^{\lambda}$. Like \cite{mer:76} in the context of jump-diffusion processes we assume that the dynamics of spikes and drops are the same in the actual and pricing measures. 
We start with finding the spot price dynamics under $\lambda$ parametrization.

Let $\lambda(u)$ be a deterministic function square-integrable on $u\in[0,T_{max}]$, where $T_{max}$ is a time horizon long enough to contain all maturities of derivatives quoted in the market,  
and introduce a new process $W_t^{\lambda}$:
\begin{equation}
W_t^{\lambda}=W_t+\int_{0}^t\frac{\lambda(u)}{\sigma_b}du,
\end{equation}
where $\sigma_b$ is the volatility of the base regime. 
From the Girsanov theorem we have that $W_t^{\lambda}$ is a Wiener process under a new measure $Q^{\lambda}$ defined as
\begin{equation}\label{eqn:pr:P_lambda}
\frac{dQ^{\lambda}}{dQ}=\exp\left[-\int_{0}^{T_{max}}\frac{\lambda(u)}{\sigma_b}dW_u-
\frac{1}{2}\int_{0}^{T_{max}}\left(\frac{\lambda(u)}{\sigma_b}\right)^2du\right]
\end{equation}
with the filtration $\mathcal{F}^W_t$, being the natural filtration of the process $W_t$.

Now, the base regime process $X_{t,b}$ can be rewritten as:
\begin{equation}\label{eqn:pr:OU:lambda}
dX_{t,b}=[\alpha-\lambda(t)-\beta X_{t,b}]dt+\sigma_bdW_t^{\lambda}
\end{equation}
and the expected future spot price is given by:
\begin{eqnarray}\label{eqn:pr:exp:market:Q}
E^{\lambda}(P_t|\mathcal{F}_0)&=&p_{bb}^{(t)}\left[X_{0}e^{-\beta t}+\frac{\alpha}{\beta}\left(1-e^{-\beta t}\right)- \int_0^t e^{-\beta(t-u)}\lambda(u)du\right]+p_{bs}^{(t)}\left(e^{\mu_s+\frac{1}{2}\sigma_s^2}+c_s\right) +\nonumber\\ &&+p_{bd}^{(t)}\left(-e^{\mu_d+\frac{1}{2}\sigma_d^2}+c_d\right)+g_t.
\end{eqnarray}

The function $\lambda(t)$ can be calibrated to the market forward prices so that $E^{\lambda}(P_t|\mathcal{F}_0)=f_0^t$, e.g.\ by using some fitting procedure (like the least squares minimization). 
Alternatively, one can find the risk premium and then use the relation between the market price of risk $\lambda(t)$ and the risk premium:
\begin{equation}\label{eqn:pr:lambda}
p_{bb}^{(t)}\int_{0}^{t}e^{-\beta (t-u)}\lambda(u)du=RP(t),
\end{equation}  
which is a simple consequence of the fact that $RP(t)=E(P_t|\mathcal{F}_0)-E^{\lambda}(P_t|\mathcal{F}_0)$,
formula (\ref{eqn:pr:exp:market:Q}) and Ito's lemma.

Now, the price of a European call option written on the electricity spot price can be derived. \\
\paragraph{Option price formula} 
If the electricity spot price $P_t$ is given by the MRS model defined by equations (\ref{eqn:pr:P})-(\ref{eqn:pr:ISLN}), then the price of a European call option written on $P_t$ with strike price $K$ and maturity $T$ is equal to:
\begin{equation}\label{eqn:pr:opt}
C_T(K)=e^{-rT}\left[p_{bb}^{(T)}C_{T,b}(K)+p_{bs}^{(T)}C_{T,s}(K)+p_{bd}^{(T)}C_{T,d}(K)\right],
\end{equation}
where
\begin{eqnarray}\label{eqn:pr:ctb}
C_{T,b}(K)=\frac{s}{\sqrt{2\pi}}\exp\left(-\frac{(K'-m)^2}{2s^2}\right)+(m-K')\left[1-\Phi\left(\frac{K'-m}{s}\right)\right],
\end{eqnarray}
\begin{eqnarray*}
C_{T,s}(K)&=&\mathbb{I}_{\{K'>c_s\}}\Bigg\{\exp\left(\mu_s+\frac{\sigma_s^2}{2}\right)\left[1-\Phi\left(\frac{\log(K'-c_s)-\mu_s-\sigma_s^2}{\sigma_s}\right)\right]-\\
&&-(K'-c_s)\left[1-F_{LN(\mu_s,\sigma^2_s)}(K'-c_s)\right]\Bigg\}+\mathbb{I}_{\{K'\leq c_s\}}\left[ \exp\left(\mu_s+\frac{\sigma_s^2}{2} \right)+c_s-K' \right]
\end{eqnarray*}
and
\begin{eqnarray*}
C_{T,d}(K)= \mathbb{I}_{\{K'< c_d\}}\Bigg\{
-\exp\left(\mu_d+\frac{\sigma_d^2}{2}\right)\Phi\left[\frac{\log(c_d-K')-\mu_d-\sigma_d^2}{\sigma_d}\right]+(c_d-K')F_{LN(\mu_d,\sigma^2_d)}(c_d-K')\Bigg\}.
\end{eqnarray*}
Further,
$K'=K-g_T$, $m=X_{0}e^{-\beta T}+\frac{\alpha}{\beta}\left(1-e^{-\beta T}\right)-\int_0^T e^{-\beta (T-u)}\lambda(u)du$, $s^2=\frac{\sigma_b^2}{2\beta}\left(1-e^{-2\beta T}\right)$ and $F_{LN(\mu,\sigma^2)}$ is the cumulative distribution function of the log-normal distribution with parameters $\mu$ and $\sigma^2$.

Note that, in order to make the exposition of the paper clear, the price derivation is moved to the Appendix.

Here, we assume that the option is settled in an infinitesimal period of time $[T,T+\Delta]$. However, in practice, the electricity spot price usually corresponds to a delivery during some period of time (e.g.\ an hour, a day) and, hence, the maturity of the option should be specified on the same time-scale. On the other hand, the analyzed spot price quotations usually represent some delivery period. For instance, if the considered data is quoted daily, as it will be in the empirical example of Section \ref{sec:ex}, then the maturity of the option would be also given in daily time-scale and would correspond to daily delivery.

\subsection{Electricity forward contracts}\label{sec:for}
Probably, the most popular electricity derivatives are the forward contracts. Recall that a forward contract is an agreement to buy (sell) a certain amount of the underlying (here MWh of electricity) at a specified future date. Settlement of the contract can be specified in two ways: with physical delivery of electricity or with only financial clearing. Both types of settlement are in the following called delivery. Denote the price at time $t$ of a forward contract with a delivery at time $T$ by $f_t^T$. Since the cost of entering a forward contract is equal to zero, the expected future payoff under the pricing measure should fulfill:
\begin{equation}
E^{\lambda}(P_T-f_t^T|\mathcal{F}_t)=0,
\end{equation}
what implies that
\begin{equation}
f_t^T=E^{\lambda}(P_T|\mathcal{F}_t).
\end{equation}
Observe, that now we define the price of a forward contract at any future date $t$. This is motivated by the fact that the valuation at time 0 of an option written on a forward contract requires the knowledge about the forward price dynamics at the option's maturity $t$.

\paragraph{Forward price formula} 
If the electricity spot price $P_t$ is given by the MRS model defined by equations (\ref{eqn:pr:P})-(\ref{eqn:pr:ISLN}), then the price at time $t$ of a forward contract written on $P_t$ with a delivery at time $T$ is given by the following formula
\begin{eqnarray} \label{eqn:pr:f:price}
f_t^T&=&P(R_{\lfloor T\rfloor}=b|\mathcal{F}_t)\left[E^{\lambda}(X_{t,b}|\mathcal{F}_t)e^{-\beta (T-t)}+\frac{\alpha}{\beta} \left(1-e^{-\beta (T-t)}\right) - \int_t^Te^{-\beta (T-u)}\lambda (u)du \right]+ \nonumber\\ && +P(R_{\lfloor T\rfloor}=s|\mathcal{F}_t)(e^{\mu_s+\frac{1}{2}\sigma_s^2}+c_s) + P(R_{\lfloor T\rfloor}=d|\mathcal{F}_t)(c_d-e^{\mu_d+\frac{1}{2}\sigma_d^2})+g_T, 
\end{eqnarray}
where $P(R_{\lfloor T\rfloor}=i|\mathcal{F}_t)=\sum_{j\in\{b,s,d\}} P(R_{\lfloor T\rfloor}=i|R_{\lfloor t\rfloor}=j)\mathbb{I}_{\{R_{\lfloor t\rfloor}=j\}}.$

Note that in the above formula $E^{\lambda}(X_{t,b}|\mathcal{F}_t)$ is used, since this expectation depends on the state process value at time $t$. Namely, if $R_t=b$ then $E^{\lambda}(X_{t,b}|\mathcal{F}_t)=X_{t,b}=X_t$. On the other hand, if at time $t$ a spike or a drop occurred then $E^{\lambda}(X_{t,b}|\mathcal{F}_t)=E^{\lambda}(X_{t,b}|\mathcal{F}_{t-1})$ and again this expectation is dependent on $R_{t-1}$ value. A general formula for $E^{\lambda}(X_{t,b}|\mathcal{F}_t)$ can be found using the same derivations as in Remark \ref{rem}.

When deriving the forward price dynamics, we have to remember that the properties of the obtained model should comply with the observed market prices. One of the most pronounced features of the market forward prices is the observed term structure of volatility, called the Samuelson effect. Precisely, the volatility of the forward prices is quite law for distant delivery periods, however, it increases rapidly with approaching maturity of the contracts. Here, the forward price volatility is described by the part  $P(R_{\lfloor T\rfloor}=b|\mathcal{F}_t)E^{\lambda}(X_{t,b}|\mathcal{F}_t)e^{-\beta (T-t)}$ of formula (\ref{eqn:pr:f:price}). Hence, it is specified by the volatility of the spot price base regime scaled with $e^{-\beta (T-t)}$ and the corresponding probability of switching to the base regime. Observe that the scaling factor $e^{-\beta (T-t)}$ exhibits the Samuelson effect as it increases to 1 with $t$ approaching maturity time $T$. Moreover the forward price volatility, again due to the scaling factor, is lower than the spot price volatility. This is in compliance with the behavior of the market spot and forward prices. 
 
Electricity forward contracts listed on energy exchanges are usually settled during a certain period of time (a week, a month, a year, etc.). Denote the price at time $t$ of a forward contract settled during the period $[T_1,T_2]$ by $f_t^{[T_1,T_2]}$. Obviously, the latter is the mean price of forward contracts with delivery during the period $[T_1,T_2]$, namely:
\begin{equation}\label{eqn:pr:forward:T1T2}
f_t^{[T_1,T_2]}=\int_{T_1}^{T_2}w(T_1,T_2,T)f_t^T dT=\int_{T_1}^{T_2}w(T_1,T_2,T)E^{\lambda}(P_T|\mathcal{F}_t) dT,
\end{equation}
where $w(T_1,T_2,T)$ is the weight function representing the time value of money. The form of $w$ depends on the contract specification. For contracts settled at maturity we have $w(T_1,T_2,T)=\frac{1}{T_2-T_1}$, while for instant settlement $w(T_1,T_2,T)=\frac{re^{-rT}}{e^{-rT_1}-e^{-rT_2}}$, where $r>0$ is the interest rate \citep{ben:ben:koe:08}.
The price $f_t^{[T_1,T_2]}$ can be obtained from formulas (\ref{eqn:pr:f:price}) and (\ref{eqn:pr:forward:T1T2}). Indeed, we have:
\begin{eqnarray}\label{eqn:pr:forward:delivery}
f_t^{[T_1,T_2]}&=&E^{\lambda}(X_{t,b}|\mathcal{F}_t)\int_{T_1}^{T_2} w(T_1,T_2,T)P(R_{\lfloor T\rfloor}=b|\mathcal{F}_t)e^{-\beta (T-t)}dT + \nonumber\\ 
&&+\int_{T_1}^{T_2}w(T_1,T_2,T)P(R_{\lfloor T\rfloor}=b|\mathcal{F}_t)\left[\frac{\alpha}{\beta}\left(1-e^{-\beta (T-t)}\right)-\int_t^Te^{-\beta(T-u)}\lambda(u)du\right]dT+ \nonumber\\ 
&& + (e^{\mu_s+\frac{1}{2}\sigma_s^2}+c_s)\int_{T_1}^{T_2} w(T_1,T_2,T)P(R_{\lfloor T\rfloor}=s|\mathcal{F}_t)dT+ \nonumber\\ 
&&  + (c_d-e^{\mu_d+\frac{1}{2}\sigma_d^2})\int_{T_1}^{T_2} w(T_1,T_2,T)P(R_{\lfloor T\rfloor}=d|\mathcal{F}_t)dT +\int_{T_1}^{T_2}w(T_1,T_2,T) g_T dT.
\end{eqnarray}

\subsection{Options written on electricity forward contracts}\label{sec:pr:op:forward}
Finally, we find an explicit formula for a European call option written on a forward contract delivering electricity during a specified period of time. Observe, that the forward price $f_t^{[T_1,T_2]}$ depends on the spot price at time $t$ and, as a consequence, also on the state process value at time $t$. 

We consider an option written on an electricity forward contract with settlement during a specified period of time, as it is the most popular specification of electricity options on energy exchanges. For example, in the EEX market there are options written on forward contracts with monthly, quarterly and yearly settlement periods. The maturity of such options is set to the fourth business day before the beginning of the underlying contract's settlement period. The EEX example will be examined in Section \ref{sec:ex}.

\paragraph{Price formula for an option written on a forward contract} 
If the electricity spot price $P_t$ is given by the model defined by equations (\ref{eqn:pr:P})-(\ref{eqn:pr:ISLN}), then the price of a European call option with strike price $K$ and maturity $t$ written on a forward contract with delivery during the period $[T_1,T_2]$ is equal to:
\begin{eqnarray}
Cf_t^{[T_1,T_2]}(K)&=&e^{-rt}\Bigg\{A_0(b)C_{t,b}\left(\frac{K-B_0(b)}{A_0(b)}+g_t\right)P(R_{\lfloor t\rfloor}=b|R_0=b)+\nonumber\\
&&+\sum_{i\in{\{s,d\}}}\sum_{k=1}^{\lfloor t\rfloor} \Bigg[A_k(i)C_{\lfloor t\rfloor-k+1,b}\left(\frac{K-B_k(i)}{A_k(i)}+g_{\lfloor t\rfloor-k+1}\right)\times\nonumber\\
&&\times P(R_{\lfloor t\rfloor}=i,R_{\lfloor t\rfloor-1}\neq b ,...,R_{\lfloor t\rfloor-k}=b|R_0=b)\Bigg]\Bigg\},
\end{eqnarray}
where 
\begin{eqnarray}\label{eqn:Aki}
A_k(i)&=&\int_{T_1}^{T_2}w(T_1,T_2,T) P(R_{\lfloor T\rfloor}=b|R_{\lfloor t\rfloor}=i) e^{-\beta (T-\lfloor t\rfloor+k-1)}dT,\\
A_0(b)&=&\int_{T_1}^{T_2}w(T_1,T_2,T) P(R_{\lfloor T\rfloor}=b|R_{\lfloor t\rfloor}=b)e^{-\beta (T-t)}dT,
\end{eqnarray}
\begin{eqnarray}\label{eqn:Bki}
B_k(i)&=&\int_{T_1}^{T_2}w(T_1,T_2,T)P(R_{\lfloor T\rfloor}=b|R_{\lfloor t\rfloor}=i) \times \nonumber \\ 
&&\times \left[\frac{\alpha}{\beta}\left(1-e^{-\beta (T-\lfloor t\rfloor+k-1)}\right)-\int_{\lfloor t\rfloor-k+1}^Te^{-\beta(T-u)}\lambda(u)du\right]dT+ \nonumber\\ 
&& + (e^{\mu_s+\frac{1}{2}\sigma_s^2}+c_s)\int_{T_1}^{T_2}w(T_1,T_2,T) P(R_{\lfloor T\rfloor}=s|R_{\lfloor t\rfloor}=i)dT + \nonumber\\ 
&& + (c_d-e^{\mu_d+\frac{1}{2}\sigma_d^2})\int_{T_1}^{T_2}w(T_1,T_2,T) P(R_{\lfloor T\rfloor}=d|R_{\lfloor t\rfloor}=i)dT+\int_{T_1}^{T_2} w(T_1,T_2,T)g_T dT,
\end{eqnarray}
\begin{eqnarray}\label{eqn:B0}
B_0(b)&=&\int_{T_1}^{T_2}w(T_1,T_2,T)P(R_{\lfloor T\rfloor}=b|R_{\lfloor t\rfloor}=b) \left[\frac{\alpha}{\beta}\left(1-e^{-\beta (T-t)}\right)-\int_{t}^Te^{-\beta(T-u)}\lambda(u)du\right]dT+ \nonumber\\ 
&& + (e^{\mu_s+\frac{1}{2}\sigma_s^2}+c_s)\int_{T_1}^{T_2}w(T_1,T_2,T) P(R_{\lfloor T\rfloor}=s|R_{\lfloor t\rfloor}=b)dT + \nonumber\\ 
&& + (c_d-e^{\mu_d+\frac{1}{2}\sigma_d^2})\int_{T_1}^{T_2}w(T_1,T_2,T) P(R_{\lfloor T\rfloor}=d|R_{\lfloor t\rfloor}=b)dT+ \int_{T_1}^{T_2} w(T_1,T_2,T)g_T dT.
\end{eqnarray}
and $C_{t,b}(K)$ is the `base regime part' of the price of a European call option written on the electricity spot price with maturity $t$ and strike $K$, see equation (\ref{eqn:pr:ctb}) with $T=t$.

\section{EEX market example}\label{sec:ex}
Theoretical results from the previous Sections allow us to price energy derivatives. We assume that the electricity spot price follows the model specified by equations (\ref{eqn:pr:P})-(\ref{eqn:pr:ISLN}) with a periodic transition matrix and $c_s$, $c_d$ being the first and third quartile of the dataset, respectively. We use mean daily EEX spot prices from the period January 2, 2006 - January 2, 2011 (5 years and 261 whole weeks). In order to calibrate the model, we first remove the seasonal component.

\begin{figure}
\centering
\includegraphics[scale=0.9]{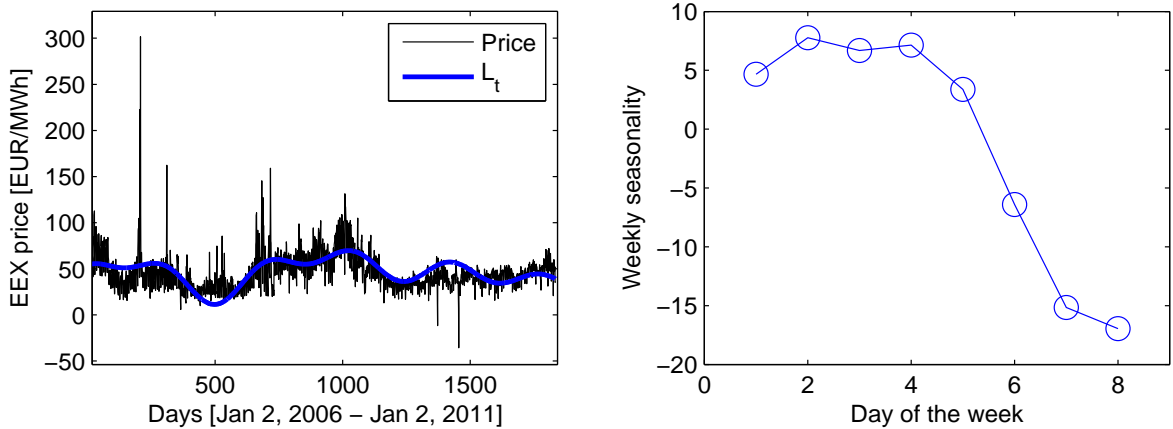}
\caption{ Left panel: EEX spot prices and the fitted long term seasonal component (blue solid line). Right panel: weekly periodicity calculated for the EEX spot prices. Note, that 1 denotes Monday, 2 -- Tuesday, ..., 7 -- Sunday, while the eight day of the week is used for the German national holidays.}
\label{fig:pr:seasonality}
\end{figure}

We assume that the deterministic function $g_t$ is composed of two parts: a long term trend $L_t$ and a weekly seasonality $S_t$. 
Since the valuation of derivatives requires forecasting the seasonal component, we model the long term trend by a sum of sine functions:
\begin{equation} \label{eqn:pr:gt}
L_t= \left(a_1 + a_2t\right)\sin\left[ 2\pi\left(t+a_3\right)\right] +  \left(a_4 + a_5t\right)\sin\left[ 2\pi a_6\left(t+a_7\right)\right] + a_8 + a_9t + a_{10}t^2,
\end{equation}
where $t$ is in yearly time scale. Note, that the first component of the above sum is responsible for the yearly periodicity, while the second one captures seasonalities of different period than one year (here, we obtain nearly half-year period, see $a_6$ in Table \ref{tab:pr:a}). The function $L_t$ is fitted to the EEX spot prices using the least squares method. The obtained curve is plotted in Figure \ref{fig:pr:seasonality}, while the obtained $a_i$ coefficients are given in Table \ref{tab:pr:a}.

\begin{table}
\centering
\small
\caption{Coefficients of the function $L_t$, see equation (\ref{eqn:pr:gt}), fitted to the EEX spot prices.}
\label{tab:pr:a}
\begin{tabular}{cccccccccc}
\hline
\multicolumn{1}{c}{$a_1$} &
\multicolumn{1}{c}{$a_2$} & \multicolumn{1}{c}{$a_3$}&
\multicolumn{1}{c}{$a_4$} & \multicolumn{1}{c}{$a_5$} &
\multicolumn{1}{c}{$a_6$} & \multicolumn{1}{c}{$a_7$} &
\multicolumn{1}{c}{$a_8$} & \multicolumn{1}{c}{$a_9$} &
\multicolumn{1}{c}{$a_{10}$} \\\hline
 -11.99 & 0.55 & -0.13 &34.03 & -8.04 &  0.46  & 6.75 &  25.37  & 19.20  & -3.35 \\
 \hline\end{tabular}
\end{table}
The estimated long term trend is subtracted from the analyzed time series. Next, the short term seasonal component is estimated using the `average week' method. Namely, we calculate the mean of prices corresponding to each day of the week (German national holidays are treated as the eight day of the week). The obtained weekly pattern is plotted in Figure \ref{fig:pr:seasonality}.

\begin{figure}
\centering
\includegraphics[scale=0.8]{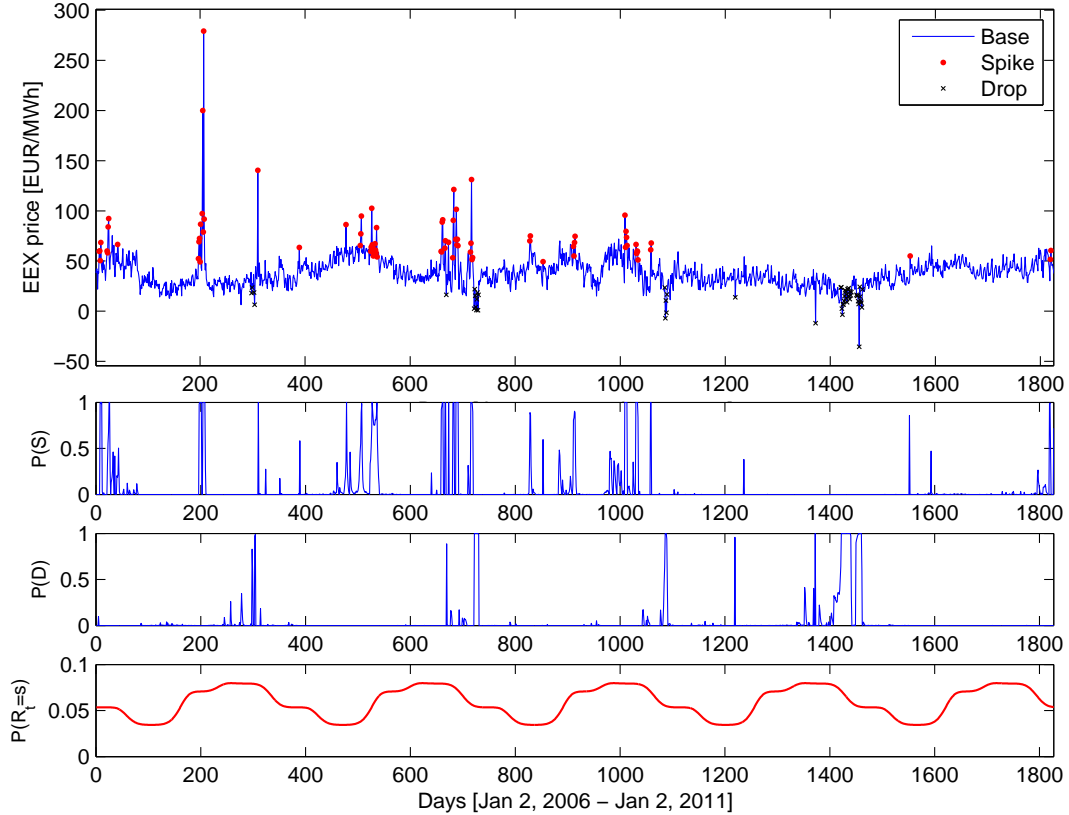}
\caption{Calibration results of the 3-regime MRS model fitted to the EEX deseasonalized prices. 
The prices classified to the spike regime i.e.\ with $P(S)=P(R_t=s|X_1,X_2,...,X_T)>0.5$ are denoted by red dots, while the pricess classified to the drop regime i.e.\ with $P(D)=P(R_t=d|X_1,X_2,...,X_T)>0.5$ are denoted with black x's. Additionally, in the lower panels the corresponding probabilities are plotted. The estimated unconditional probability of spike occurrence is displayed in the bottom panel. Observe, that the highest probability of spike is obtained for the Autumn/Winter period, while the lowest for Spring.}
\label{fig:pr:eex}
\end{figure}

Finally, the deseasonalized prices are obtained by subtracting the long and short term trend from the EEX spot prices. Moreover, the data is shifted so that the minimum of the deseasonalized and the original prices is the same. The resulting time series can be seen in Figure \ref{fig:pr:eex}.

After removing seasonality, we are left with modeling the stochastic part $X_t$. Here, we calibrate the 3-regime MRS model, see equations (\ref{eqn:model_def})-(\ref{eqn:pr:ISLN}), to the deseasonalized EEX prices. To this end, we use a version of the Expectation-Maximization algorithm of \cite{dem:lai:rub:77}, which was applied to MRS models by \cite{ham:90} and was later refined by \cite{kim:94}. For the detailed description of the algorithm in case of the 3-regime model considered in this paper see the recent work of \cite{jan:wer:12:est}. The obtained parameters are given in Table \ref{tab:pr:param}. Observe high probabilities of staying in the same regime, ranging from 0.40 for the drop regime up to 0.97 for the base regime. Hence, the assumed model allows for modeling consecutive spikes or drops in a very natural way. The calibration results are plotted in Figure \ref{fig:pr:eex}, where additionally the regimes classification is illustrated. As we may observe spikes/drops occur usually as a series of high/low prices rather than separate outstanding observations. What is interesting to note, is the clear seasonal pattern in the estimated probability of spike, see the bottom panel in Figure \ref{fig:pr:eex}. Indeed, the highest spike occurrence probability is obtained for the Autumn/Winter season, while the lowest for Spring.

\begin{table}
\centering
\small
\caption{Parameter estimates of the 3-regime MRS model fitted to the deseasonalized EEX prices.}
\label{tab:pr:param}
\begin{tabular}{cccccccccc}
\hline
\multicolumn{7}{c}{Parameters} &\multicolumn{3}{c}{Probabilities} \\
\hline
\multicolumn{1}{c}{$\alpha_1$} &
\multicolumn{1}{c}{$\beta_1$} & \multicolumn{1}{c}{$\sigma^2_1$}&
\multicolumn{1}{c}{$\alpha_2$} & \multicolumn{1}{c}{$\sigma^2_2$} &
\multicolumn{1}{c}{$\alpha_3$} & \multicolumn{1}{c}{$\sigma^2_3$} &
\multicolumn{1}{c}{$p_{11}$} & \multicolumn{1}{c}{$p_{22}$} &
\multicolumn{1}{c}{$p_{33}$} \\\hline
 5.98 & 0.16  & 39.53 &2.89 & 0.64 &  2.62 & 0.33 &  0.97 & 0.66 & 0.40 \\
 \hline\end{tabular}
\end{table}

In order to validate the used MRS model, we apply a Kolmogorov-Smirnov goodness-of-fit test for the marginal distribution of the individual regimes, as well as, for the whole model. We use two testing procedures. The first one (called ewedf -- equally weighted empirical distribution function) is based on classifying observations to the most probable regime, i.e.\ assuming that $R_t=i$ if $P(R_t=i|X_1,X_2,...,X_T)>0.5$. As a consequence, the standard Kolmogorov-Smirnov goodness-of-fit test can be applied. The second one (called wedf) utilizes a notion of the weighted empirical distribution function, where $t$-th observation is taken into account with weight proportional to the probability $P(R_t=i|X_1,X_2,...,X_T)$. For the detailed testing procedure derivation see \cite{jan:wer:12:gof}. The obtained test $p$-values are given in Table \ref{tab:pr:tests}. Recall, that $p$-value higher than 5\% means that we cannot reject, at the 5\% significance level, the hypothesis that the analyzed dataset was driven by the assumed model. As all of the obtained $p$-values are higher than 5\%, we cannot reject the considered 3-regime MRS model as a proper one for the analyzed dataset.

\begin{table}
\centering
\small
\caption{$p$-values of the goodness-of-fit tests for the 3-regime MRS model fitted to the deseasonalized EEX prices. The test results for the ewedf, as well as, the wedf approach are provided.}
\label{tab:pr:tests}
\begin{tabular}{ccccccccc}
\hline
&\multicolumn{4}{c}{ewedf} &\multicolumn{4}{c}{wedf} \\
\hline
Regime&\multicolumn{1}{c}{Base} &
\multicolumn{1}{c}{Spike} & \multicolumn{1}{c}{Drop}&
\multicolumn{1}{c}{Model} &\multicolumn{1}{c}{Base} &
\multicolumn{1}{c}{Spike} & \multicolumn{1}{c}{Drop}&
\multicolumn{1}{c}{Model} \\\hline
$p$-value & 0.64 & 0.16 & 0.50 & 0.34 & 0.30 & 0.59 & 0.95& 0.25 \\
 \hline\end{tabular}
\end{table}

Before we start with the valuation of derivatives we have to find the risk premium and the function $\lambda$, see equation (\ref{eqn:pr:lambda}). We use monthly forward contracts listed on the EEX market on January 3, 2011, i.e.\ on the day directly following the calibration period. The prices, as well as, the delivery periods of the analyzed forward contracts are given in Table \ref{tab:pr:f}.
\begin{table}
\centering
\small
\caption{Specification of the monthly forward contracts listed on the EEX market on January 3, 2011.}
\label{tab:pr:f}
\begin{tabular}{rccc}
\hline
\multicolumn{1}{c}{Name}&	Settlement Price& $T_1$&  $T_2$	\\	\hline
Feb-11&	54.35	&1.2.2011	&28.2.2011\\
Mar-11&	51.64	&1.3.2011	&31.3.2011\\
Apr-11&	48.07 &1.4.2011   &30.4.2011\\
May-11&	45.53	&1.5.2011	&31.5.2011\\
Jun-11&	48.50	&1.6.2011	&30.6.2011\\
Jul-11&	49.00	&1.7.2011	&31.7.2011\\ \hline\end{tabular}
\end{table}

\begin{figure}
\centering
\includegraphics[scale=0.8]{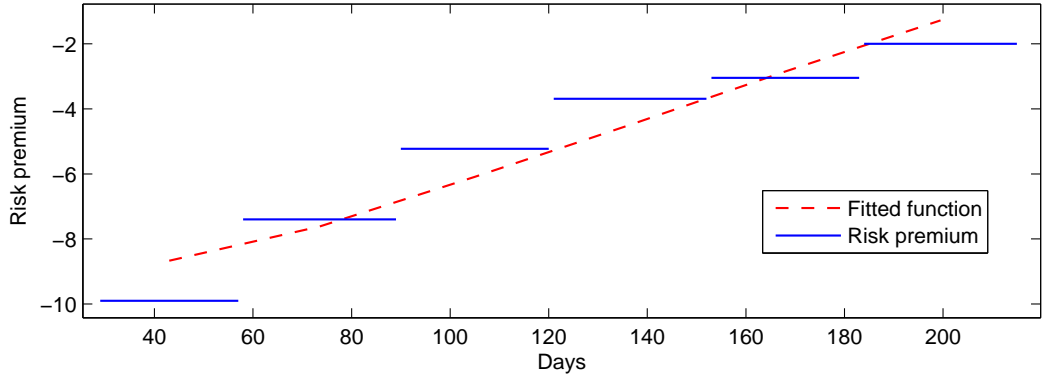}
\caption{The risk premium obtained from monthly forward contracts with different delivery periods (blue lines), as well as, the fitted function (see equation (\ref{eqn:pr:lambdaT})) plotted with red dashed line. }
\label{fig:pr:rp}
\end{figure}

Assume that $\lambda(t)=\lambda_1 t+\lambda_2$. Since we analyze monthly contracts, the risk premium should be also calculated on the monthly basis, i.e.\ instead of equation (\ref{eqn:pr:rp}) we use:
\begin{equation}
RP(T_1,T_2)=\frac{1}{T_2-T_1+1}\sum_{t=T_1}^{T_2} E(P_t|\mathcal{F}_0)-f_{0}^{[T_1,T_2]}.
\end{equation}
Note that the average monthly expected spot price is calculated as an arithmetic mean instead of an integral, because the analyzed spot prices are quoted is discrete time (on a daily basis).
For the same reason $\lambda_1$ and $\lambda_2$ are found by fitting:
\begin{equation}\label{eqn:pr:lambdaT}
\frac{\lambda_1\sum_{t=T_1}^{T_2} p_{bb}^{(t)}(t-(1-e^{-\beta t})/\beta) +\lambda_2\sum_{t=T_1}^{T_2} p_{bb}^{(t)}(1-e^{-\beta t})}{\beta(T_2-T_1+1)}=RP(T_1,T_2).
\end{equation}
See equation (\ref{eqn:pr:lambda}) for the comparison with the continuous time scale.   
The values of the risk premium obtained from contracts with different delivery periods are plotted in Figure \ref{fig:pr:rp} (for contract specifications see Table \ref{tab:pr:f}). Observe a strong evidence for the negative risk premium, especially for contracts with approaching delivery period. For the contracts with more distant delivery period the obtained risk premium is less significant. Using the least-squares minimization scheme we get $\lambda=0.0084t-1.8387$, see the red dashed line in Figure \ref{fig:pr:rp} for the plot of the function fitted to the risk premium (i.e.\ the left hand side of equation (\ref{eqn:pr:lambdaT})).

\begin{figure}
\centering
\includegraphics[scale=0.8]{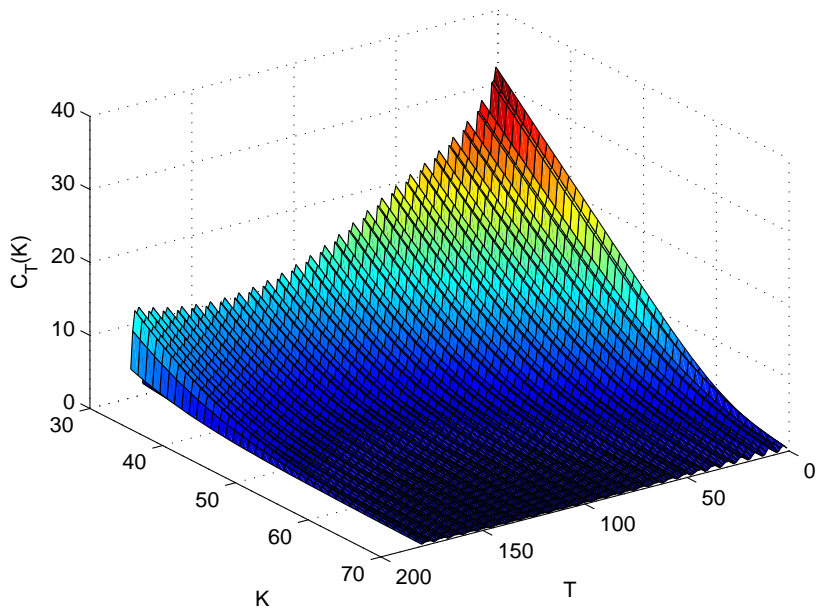}
\caption{Prices of a European call option written on the electricity spot price for different maturities $T$ and strike prices $K$. The prices are calculated on January 3, 2011.}
\label{fig:pr:C}
\end{figure}

Now, we can derive the price of a European call option written on the electricity spot price. Assume that the interest rate $r$ is equal to 0. The option prices obtained in Section \ref{sec:pr:spot} with different maturities $t$ and strike prices $K$ are plotted in Figure \ref{fig:pr:C}.
Obviously, the lower is the strike price, the higher is the call option price. What is interesting to note, is how the option price depends on the maturity tenor $t$. Observe the clear seasonal pattern of option prices, both on the weekly and the long-term level. The long-term seasonality is caused not only by the deterministic component $g_t$ but also by the periodic transition matrix allowing for varying spike (drop) probabilities during the year. Recall that in the EEX market the spike probability is the highest in Autumn/Winter and the lowest in Spring, see Figure \ref{fig:pr:eex}.

Now, we derive the prices of European call options written on monthly forward contracts. Using the results obtained in Section \ref{sec:pr:op:forward} we calculate the price of an option written on the forward contract with settlement in February 2011. The results are plotted in Figure \ref{fig:pr:eex}.
\begin{figure}
\centering
\includegraphics[scale=0.8]{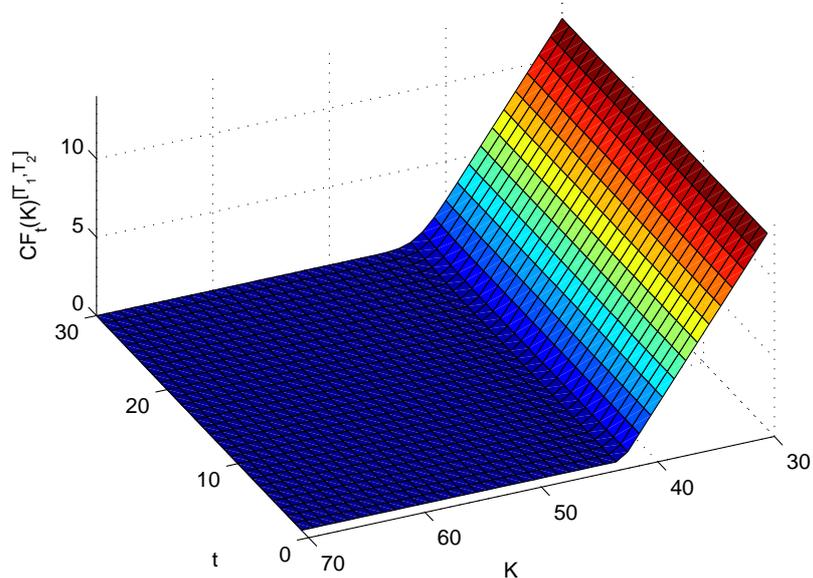}
\caption{Prices of a European call option written on the electricity forward contract with settlement in February 2011 for different maturities $t$ and strike prices $K$. The prices are
 calculated on January 3, 2011.}
\label{fig:pr:CF}
\end{figure}
In order to check how the option price varies according to the delivery period, we calculate the prices of options written on monthly forward contracts with deliveries within the next 6 months (i.e.\ February 2011 till July 2011). According to the products specification in the EEX market, the maturity of the options is set to the fourth business day before the beginning of the delivery period. The obtained option prices are given in Table \ref{tab:pr:CF}.
\begin{table}
\centering
\small
\caption{Prices of European call options written on monthly forward contracts with strike price $K=30$. The prices are calculated on January 3, 2011.}
\label{tab:pr:CF}
\begin{tabular}{lcc}
\hline
\multicolumn{1}{c}{Month of delivery $[T_1,T_2]$} & Options maturity $t$ & Options price $Cf_t^{[T_1,T_2]}(K)$\\
\hline
February &26.1.2011&13.5449\\
March&23.2.2011&9.1074\\
April&28.3.2011&4.2258\\
May&26.4.2011&2.6826\\
June&28.5.2011& 3.6148\\
July&27.6.2011&7.1913\\
\hline
\end{tabular}
\end{table}
Similarly, as in the case of options written on the spot price, we observe the lowest option prices for settlement during the Spring months.

\section{Conclusions}\label{sec:con}
In this paper we have derived premiums of European options written on electricity spot, as well as, forward prices. We assumed that electricity spot prices can be described by a 3-regime MRS model with independent spikes and drops and periodic transition matrix, proposed earlier by \cite{jan:wer:10:ee}.
The forward prices were then derived using the risk premium approach and fitting the model-based prices to the observed forward curve. Next, using the spot and forward price dynamics we calculated prices of the corresponding European options.

The assumed model was then calibrated to the spot prices from the European Energy Exchange. We have validated the model choice by performing a statistical goodness-of-fit test. Next, using monthly forward contracts quotations we have calculated the risk premium. We have obtained negative values, especially significant for contracts with approaching maturity. For contracts with distant settlement the risk premium values were higher. 

Finally, the presented methodology and the calibration results allowed us to find prices of European options currently listed on the EEX market. As the assumed 3-regime MRS model seems to be adequate to describe dynamics of electricity spot prices, the results of the paper can be used for reasonable pricing of electricity derivatives and, hence, yield an effective risk management tool.

\section*{Acknowledgments}
We thank Tomasz Piesiewicz from Tauron PE for the EEX spot price data. This work was supported by funds from the National Science Centre (NCN) through grant no. 2011/01/B/HS4/01077.

\section*{Appendix}
\paragraph{Derivation of the option price formula}
Using standard arguments the option price is the discounted expected value of the payoff function under the pricing measure \citep[see e.g.\ ][]{mus:rut:97}. Moreover, analogously to formula (\ref{eqn:pr:exp}) this expectation is equal to:
\begin{eqnarray}\label{eqn:prCt}
C_T(K)&=&e^{-rT}E^{\lambda}\left[(P_T-K)^+|\mathcal{F}_0\right]=e^{-rT}E^{\lambda}\left[(X_T+g_T-K)^+|\mathcal{F}_0\right]=\\\nonumber
&=&e^{-rT}\Bigg\{p_{bb}^{(T)}E^{\lambda}\left[(X_{T,b}-K')^+|\mathcal{F}_0\right]+p_{bs}^{(T)}E\left[(X_{\lfloor T\rfloor,s}-K')^+|\mathcal{F}_0\right]+p_{bd}^{(T)}E\left[(X_{\lfloor T\rfloor,d}-K')^+|\mathcal{F}_0\right]\Bigg\},
\end{eqnarray}
where $K'=K-g_T$ and $(x)^+=\max(0,x)$.\\

We start with pricing the base regime part $C_{T,b}(K)$: 
\begin{equation}
C_{T,b}(K)=E^{\lambda}\left[(X_{T,b}-K')^+|\mathcal{F}_0\right]=\int_{K'}^{\infty}(x-K')f_{X_{T,b}|X_{0,b}}(x)dx,
\end{equation}
where $f_{X_{T,b}|X_{0,b}}(x)$ denotes the density of $X_{T,b}$ conditional on $X_{0,b}$.
From (\ref{eqn:pr:OU:lambda}) and Ito's lemma we have that:
\begin{equation}
X_{T,b}=X_{0,b}e^{-\beta T}+\frac{\alpha}{\beta}\left(1-e^{-\beta T}\right)-\int_0^T e^{-\beta (T-u)}\lambda(u)du+\sigma_b\int_0^T e^{-\beta (T-u)}dW_u^{\lambda}.
\end{equation}
Hence, $X_{T,b}$ given $X_{0,b}$ has a Gaussian distribution with mean
\begin{equation}
E(X_{T,b}|\mathcal{F}_0)=X_{0,b}e^{-\beta T}+\frac{\alpha}{\beta}\left(1-e^{-\beta T}\right)-\int_0^T e^{-\beta (T-u)}\lambda(u)du,
\end{equation}
and variance
\begin{equation}
Var(X_{T,b}|\mathcal{F}_0)=\frac{\sigma_b^2}{2\beta}\left(1-e^{-2\beta T}\right).
\end{equation}
Denote the mean by $m$ and the variance by $s^2$. We have:
\begin{eqnarray}\label{eqn:pr:Ct1a}
C_{T,b}(K)&=&\frac{1}{\sqrt{2\pi}s}\int_{K'}^{\infty}(x-K')\exp\left(-\frac{(x-m)^2}{2s^2}\right)dx=\nonumber\\
&=&\frac{s}{\sqrt{2\pi}}\exp\left(-\frac{(K'-m)^2}{2s^2}\right)+(m-K')\left[1-\Phi\left(\frac{K'-m}{s}\right)\right].
\end{eqnarray}

Now, we turn to the pricing of the spike regime part.
Observe that if $K'\leq c_s$, then $E[(X_{\lfloor T\rfloor,s}-K')^+|\mathcal{F}_0]=E(X_{\lfloor T\rfloor,s})-K'$. Assume that $K'> c_s$. Denote the log-normal pdf by $f_{LN(\mu_s,\sigma^2_s)}$ and the cdf by $F_{LN(\mu_s,\sigma^2_s)}$. Then
\begin{eqnarray}\label{eqn:pr:e1}
E[(X_{\lfloor T\rfloor,s}-K')^+|\mathcal{F}_0]&=& \int_{K'}^{\infty}(x-K')f_{LN(\mu_s,\sigma^2_s)}(x-c_s)dx=\nonumber\\ 
&=&\exp\left(\mu_s+\frac{\sigma_s^2}{2}\right)\left\{1-\Phi\left[\frac{\log(K'-c_s)-\mu_s-\sigma_s^2}{\sigma_s}\right]\right\}-\nonumber\\
&&-(K'-c_s)\left[1-F_{LN(\mu_s,\sigma^2_s)}(K'-c_s)\right]
\end{eqnarray}
and we have that
\begin{eqnarray}\label{eqn:pr:Ct2}
E[(X_{\lfloor T\rfloor,s}-K')^+|\mathcal{F}_0]&=&
\mathbb{I}_{\{K'> c_s\}}\Bigg\{\exp\left(\mu_s+\frac{\sigma_s^2}{2}\right)\left[1-\Phi\left(\frac{\log(K'-c_s)-\mu_s-\sigma_s^2}{\sigma_s}\right)\right]-\nonumber\\
&&  -(K'-c_s)\left[1-F_{LN(\mu_s,\sigma^2_s)}(K'-c_s)\right]\Bigg\}+\nonumber\\
&&+\mathbb{I}_{\{K'\leq c_s\}}\left[\exp\left(\mu_s+\frac{\sigma_s^2}{2}\right)+c_s-K'\right].
\end{eqnarray}
Similarly, we can price the drop regime part:
\begin{eqnarray}\label{eqn:pr:Ct3}
E[(X_{\lfloor T\rfloor,d}-K')^+|\mathcal{F}_0]&=& \mathbb{I}_{\{K'< c_d\}}\Bigg\{
-\exp\left(\mu_d+\frac{\sigma_d^2}{2}\right)\Phi\left[\frac{\log(c_d-K')-\mu_d-\sigma_d^2}{\sigma_d}\right]+\nonumber\\
&& \qquad \qquad+(c_d-K')F_{LN(\mu_d,\sigma^2_d)}(c_d-K')\Bigg\}.
\end{eqnarray}
Finally, letting $C_{T,i}(K)=E[(X_{\lfloor T\rfloor,i}-K')^+|\mathcal{F}_0]$ for $i\in \{s,d\}$ and combining formulas (\ref{eqn:prCt}), (\ref{eqn:pr:Ct1a}), 
(\ref{eqn:pr:Ct2}) and (\ref{eqn:pr:Ct3}) yields the result.

\paragraph{Derivation of the forward price}
First, note that
\begin{eqnarray} \label{eqn:pr:forward}
E^{\lambda}(P_T|\mathcal{F}_t)=E^{\lambda}(X_T|\mathcal{F}_t)+g_T=E^{\lambda}(\mathbb{I}_{\{R_{\lfloor T\rfloor}=b\}}X_{T,b}+\mathbb{I}_{\{R_{\lfloor T\rfloor}=s\}}X_{\lfloor T\rfloor,s}+\mathbb{I}_{\{R_{\lfloor T\rfloor}=d\}}X_{\lfloor T\rfloor,d}|\mathcal{F}_t)+g_T.
\end{eqnarray}
Since $X_{\lfloor T\rfloor,i}$, for $i=s,d$, is independent of $R_{\lfloor T\rfloor}$ and $\mathcal{F}_t$, we have:
$E^{\lambda}(\mathbb{I}_{\{R_{\lfloor T\rfloor}=i\}}X_{T,i}|\mathcal{F}_t)=P(R_{\lfloor T\rfloor}=i|\mathcal{F}_t)E(X_{\lfloor T\rfloor,i})$, $i=s,d$. Secondly, we have $$E^{\lambda}(\mathbb{I}_{\{R_{\lfloor T\rfloor}=b\}}X_{T,b}|\mathcal{F}_t)=E^{\lambda}(\mathbb{I}_{\{R_{\lfloor T\rfloor}=b\}}|\mathcal{F}_t) E^{\lambda}(X_{T,b}|\mathcal{F}_t)=P(R_{\lfloor T\rfloor}=b|\mathcal{F}_t)E^{\lambda}\left(X_{T,b}|\mathcal{F}_t\right)$$ and 
from the base regime definition and Ito's lemma:
\begin{eqnarray}\label{eqn:pr:forward:base}
E^{\lambda}(X_{T,b}|\mathcal{F}_t)&=&E^{\lambda}(X_{t,b}|\mathcal{F}_t)e^{-\beta (T-t)}+\frac{\alpha}{\beta}\left(1-e^{-\beta(T- t)}\right)-\int_{t}^{T}e^{-\beta (T-u)}\lambda(u)du.\nonumber\\
\end{eqnarray}

Moreover, from the definitions of the spike and drop regimes, see equations (\ref{eqn:pr:SLN}) and (\ref{eqn:pr:ISLN}), we get:
\begin{equation}\label{eqn:pr:forward:spike}
E(X_{\lfloor T\rfloor,s})=e^{\mu_s+\frac{1}{2}\sigma_s^2}+c_s
\end{equation}
and
\begin{equation}\label{eqn:pr:forward:drop}
E(X_{\lfloor T\rfloor,d})=c_d-e^{\mu_d+\frac{1}{2}\sigma_d^2}.
\end{equation}
Finally, combining formula (\ref{eqn:pr:forward}) with (\ref{eqn:pr:forward:base}), (\ref{eqn:pr:forward:spike}) and (\ref{eqn:pr:forward:drop}) yields the result.

\paragraph{Derivation of the price formula for an option written on a forward contract}
We start the derivation of the option price formula with the following observation.

If $R_{\lfloor t\rfloor}=b$, then the forward price is given by:
\begin{eqnarray}\label{eqn:pr:lem:a}
f_{t|\{R_{\lfloor t\rfloor}=b\}}^{[T_1,T_2]} &=&X_{t,b}\int_{T_1}^{T_2}w(T_1,T_2,T) P(R_{\lfloor T\rfloor}=b|R_{\lfloor t\rfloor}=b)e^{-\beta (T-t)}dT + \nonumber\\ 
&&+\int_{T_1}^{T_2}w(T_1,T_2,T)P(R_{\lfloor T\rfloor}=b|R_{\lfloor t\rfloor}=b) \left[\frac{\alpha}{\beta}\left(1-e^{-\beta (T-t)}\right)-\int_{t}^Te^{-\beta(T-u)}\lambda(u)du\right]dT+ \nonumber\\ 
&& + (e^{\mu_s+\frac{1}{2}\sigma_s^2}+c_s)\int_{T_1}^{T_2}w(T_1,T_2,T) P(R_{\lfloor T\rfloor}=s|R_{\lfloor t\rfloor}=b)dT + \\\nonumber 
&& + (c_d-e^{\mu_d+\frac{1}{2}\sigma_d^2})\int_{T_1}^{T_2}w(T_1,T_2,T) P(R_{\lfloor T\rfloor}=d|R_{\lfloor t\rfloor}=b)dT+ \int_{T_1}^{T_2} w(T_1,T_2,T)g_T dT.
\end{eqnarray}
Moreover, if $k$ is such a number that $R_{\lfloor t\rfloor}=i, R_{\lfloor t\rfloor-1}\neq b,..., R_{\lfloor t\rfloor-k+1}\neq b, R_{\lfloor t\rfloor-k}=b$, for $i\in\{s,d\}$ (i.e.\ the last base regime price before time $t$ was observed $k$-periods earlier), then the forward price is given by: 
\begin{eqnarray}\label{eqn:pr:lem:b}
f_{t|\{R_{\lfloor t\rfloor}=i, R_{\lfloor t\rfloor-1}\neq b,..., R_{\lfloor t\rfloor-k}=b\}}^{[T_1,T_2]} &=&X_{\lfloor t\rfloor-k+1,b}\int_{T_1}^{T_2}w(T_1,T_2,T) P(R_{\lfloor T\rfloor}=b|R_{\lfloor t\rfloor}=i) e^{-\beta (T-\lfloor t\rfloor+k-1)}dT + \nonumber\\
&&+\int_{T_1}^{T_2}w(T_1,T_2,T)P(R_{\lfloor T\rfloor}=b|R_{\lfloor t\rfloor}=i) \times \nonumber \\ 
&&\times \left[\frac{\alpha}{\beta}\left(1-e^{-\beta (T-\lfloor t\rfloor+k-1)}\right)-\int_{\lfloor t\rfloor-k+1}^Te^{-\beta(T-u)}\lambda(u)du\right]dT+ \nonumber\\ 
&& + (e^{\mu_s+\frac{1}{2}\sigma_s^2}+c_s)\int_{T_1}^{T_2}w(T_1,T_2,T) P(R_{\lfloor T\rfloor}=s|R_{\lfloor t\rfloor}=i)dT + \nonumber\\ 
&& + (c_d-e^{\mu_d+\frac{1}{2}\sigma_d^2})\int_{T_1}^{T_2}w(T_1,T_2,T) P(R_{\lfloor T\rfloor}=d|R_{\lfloor t\rfloor}=i)dT+ \nonumber\\
&& + \int_{T_1}^{T_2} w(T_1,T_2,T)g_T dT.
\end{eqnarray}

Formula (\ref{eqn:pr:lem:a}) is a simple consequence of equation (\ref{eqn:pr:forward:delivery}) and the fact that $R_{\lfloor t\rfloor}=b$ implies that $E^{\lambda}(X_{t,b}|\mathcal{F}_t)=X_{t,b}$. 
In order to show (\ref{eqn:pr:lem:b}), observe that for  $R_{\lfloor t\rfloor}=i, R_{\lfloor t\rfloor-1]}\neq b,..., R_{\lfloor t\rfloor-k}=b$, $i=s,d$, we have
\begin{eqnarray}\label{eqn:pr:forward:base:exp}
E^{\lambda}(X_{t,b}|\mathcal{F}_{t})&=&E^{\lambda}\Big[X_{\lfloor t\rfloor-k+1,b}e^{-\beta(t-\lfloor t\rfloor+k-1)}+\frac{\alpha}{\beta}\left(1-e^{-\beta (t-\lfloor t\rfloor+k-1)}\right)-\nonumber\\
&&-\int_{\lfloor t\rfloor-k+1}^te^{-\beta(t-u)}\lambda(u)du+\sigma_b \int_{\lfloor t\rfloor-k+1}^t e^{-\beta (t-u)}dW_u^{\lambda}|X_{\lfloor t\rfloor-k+1,b}\Big]=\nonumber\\
&=&X_{\lfloor t\rfloor-k+1,b}e^{-\beta (t-\lfloor t\rfloor+k-1)}+\frac{\alpha}{\beta}\left(1-e^{-\beta(t-\lfloor t\rfloor+k-1)}\right)-\int_{\lfloor t\rfloor-k+1}^te^{-\beta(t-u)}\lambda(u)du.
\end{eqnarray}
Combining equations (\ref{eqn:pr:forward:delivery}) and (\ref{eqn:pr:forward:base:exp}) yields the result.

Now we can derive the option price formula.
The option price is equal to the expected future payoff. Therefore, we have 
\begin{eqnarray}\label{eqn:pr:forward:Ct}
Cf_t^{[T_1,T_2]}(K)&=&e^{-rt} E^{\lambda}\left[(f_t^{[T_1,T_2]}-K)^+|\mathcal{F}_0\right].
\end{eqnarray}
Observe that the forward price can be written as
\begin{eqnarray*}
f_t^{[T_1,T_2]}=\mathbb{I}_{\{R_{\lfloor t\rfloor}=b\}}f_{t|\{R_{\lfloor t\rfloor}=b\}}^{[T_1,T_2]}+\sum_{i\in\{s,d\}}\sum_{k=1}^{\lfloor t\rfloor}\mathbb{I}_{\{R_{\lfloor t\rfloor}=i, R_{\lfloor t\rfloor-1}\neq b,..., R_{\lfloor t\rfloor-k}=b\}}f_{t|\{R_{\lfloor t\rfloor}=i, R_{\lfloor t\rfloor-1}\neq b,..., R_{\lfloor t\rfloor-k}=b\}}^{[T_1,T_2]},
\end{eqnarray*}
where $f_{t|\{R_{\lfloor t\rfloor}=b\}}^{[T_1,T_2]}$ and $f_{t|\{R_{\lfloor t\rfloor}=i, R_{\lfloor t\rfloor-1}\neq b,..., R_{\lfloor t\rfloor-k}=b\}}^{[T_1,T_2]}$, $i\in\{s,d\}$, $k=1,2,...,\lfloor t\rfloor$, are given by (\ref{eqn:pr:lem:a}) and (\ref{eqn:pr:lem:b}). 
Hence,
\begin{eqnarray}
Cf_t^{[T_1,T_2]}(K)&=&e^{-rt} \Bigg\{E^{\lambda}\left[(f_{t|\{R_{\lfloor t\rfloor}=b\}}^{[T_1,T_2]}-K)^+|\mathcal{F}_0\right]P(R_{\lfloor t\rfloor}=b|R_0=b)+\nonumber\\
&&+\sum_{i\in\{s,d\}}\sum_{k=1}^{\lfloor t\rfloor}\Big\{ E^{\lambda}\left[(f_{t|\{R_{\lfloor t\rfloor}=i,R_{\lfloor t\rfloor-1}\neq b ,...,R_{\lfloor t\rfloor-k}=b\}}^{[T_1,T_2]}-K)^+|\mathcal{F}_0\right]\times\nonumber\\
&&\times P(R_{\lfloor t\rfloor}=i,R_{\lfloor t\rfloor-1}\neq b ,...,R_{\lfloor t\rfloor-k}=b|R_0=b)\Big\}.
\end{eqnarray}

Now, observe that
\begin{eqnarray}\label{eqn:pr:forward:Ct1} 
E^{\lambda}\left[(f_{t|\{R_{\lfloor t\rfloor}=b\}}^{[T_1,T_2]}-K)^+|\mathcal{F}_0\right]&=&E^{\lambda}\left[\left(X_{t,b}A_0(b)+B_0(b)-K\right)^+|\mathcal{F}_0\right]=\\
&=& A_0(b)E^{\lambda}\left[\left(X_{t,b}-\frac{K-B_0(b)}{A_0(b)}\right)^+\Big|\mathcal{F}_0\right]= A_0(b)C_{t,b}\left(\frac{K-B_0(b)}{A_0(b)}+g_t\right),\nonumber
\end{eqnarray}
where $C_{t,b}\left(\frac{K-B_0(b)}{A_0(b)}+g_t\right)$ is the `base regime part' of the price of a European call option written on the electricity spot price with maturity $t$ and strike $\frac{K-B_0(b)}{A_0(b)}+g_t$, see equation (\ref{eqn:pr:ctb}) and $A_k$, $B_k$ are defined in equations (\ref{eqn:Aki})-(\ref{eqn:B0}). 
Similarly,
\begin{eqnarray}\label{eqn:pr:forward:Ct2}
E^{\lambda}\left[(f_{t|\{R_{\lfloor t\rfloor}=i,R_{\lfloor t\rfloor-1}\neq b ,...,R_{\lfloor t\rfloor-k}=b\}}^{[T_1,T_2]}-K)^+|\mathcal{F}_0\right]&=& A_k(i)E^{\lambda}\left[\left(X_{\lfloor t\rfloor-k+1,b}-\frac{K-B_k(i)}{A_k(i)}\right)^+|\mathcal{F}_0\right]=\nonumber\\
&=&A_k(i)C_{\lfloor t\rfloor-k+1,b}\left(\frac{K-B_k(i)}{A_k(i)}+g_{\lfloor t\rfloor-k+1}\right)
\end{eqnarray}
for $i\in\{s,d\}$
Finally, combining formulas (\ref{eqn:pr:forward:Ct}), (\ref{eqn:pr:forward:Ct1}) and (\ref{eqn:pr:forward:Ct2}) 
completes the proof.\\


\bibliographystyle{elsarticle-harv}

\end{document}